\documentclass[twocolumn]{article}
\usepackage[dvips]{graphicx}
\begin{document}
\title{\large \bf Search Efficiency in Indexing Structures for Similarity Searching}
\author{Girish Motwani \\ Sandhya G\\Department of Computer Science and
Automation\\Indian Institute of Science}
\date{November 24, 2003}
\maketitle
\thispagestyle{empty}
\bibliographystyle{unsrt}
\begin{abstract}
Similarity searching finds application in a wide variety of domains
including multilingual databases, computational biology, pattern recognition
and text retrieval. Similarity is measured in terms of a distance function
({\it edit distance}) in general metric spaces, which is expensive to compute.
Indexing techniques can be used reduce the number of
distance computations. We present an analysis of various existing similarity
indexing structures for the same. The performance obtained using the index
structures studied was found to be unsatisfactory . We propose an indexing
technique that combines the features of clustering with M tree(MTB) and the
results indicate that this gives better performance. 
\end{abstract}
\section{Introduction}
With the advent of new application domains such as
multilingual databases, computational biology, text retrieval, pattern
recognition and function approximation, there is a need for proximity
searching, that is, searching for elements similar to a given query element.
Similarity is modeled using a distance function; this distance function along
with a set of objects defines a metric space. Computing distance function can
be expensive, for example, one of the requirements in multilingual database
systems is to 
find similar strings, where the distance({\it edit distance}) between the strings is computed using
an O(mn) algorithm where m, n are the length of the
strings compared. This necessitates the use of an efficient indexing technique which would
result in fewer distance computations at query time. Having an indexing
structure serves the dual purpose of decreasing both CPU and I/O costs.
Existing index structures such as B+ trees used in exact matching proves
inadequate for the above requirements.\\ \\
Various indexing structures have been proposed for similarity searching in metric
spaces. We present the performance analysis of these structures in terms of the
percentage of database scanned by varying edit distances from 10\% to 100\%. \\
After providing a preliminary background in Section 2, we move on to the
description of the existing index structures in Section 3. Section 4 describes
the experimental set up and the analysis is presented in Section 5. Section 6
concludes the paper. 
\section{Preliminaries}
A metric space comprises of a collection of objects and an assosciated distance
function satisfying the following properties. 
\begin{itemize}
\item {Symmetry \\ 
$d(a, b) = d(b, a)$
}
\item {Non-negativity \\
$d(a, b) > 0$ if $(a \neq b)$ and $d(a, b) = 0$ if $(a = b)$
}
\item {Triangle inequaltiy \\
$d(a, b) \leq d(a, c) + d(c, b)$
}
\end{itemize}
a, b, c are objects of the metric space. \\ \\
Edit distance({\it
Levenshtein distance}) satisfies the above mentioned properties. The edit distance between two strings is defined as the total number of simple edit operations such
as additions, deletions and substitutions required to transform one string to
another. For example, consider
the strings {\it paris} and {\it spire}. The edit distance between these two
strings is 4,
as the transformation of {\it paris} to {\it spire} requires one addition, one deletion and
two substitutions. Edit distance computation is expensive since the
alogorithmic complexity is O(mn) where m, n are the length of the strings
compared. \\

One of the common queries in applications requiring similarity search is to
find all elements within a given edit distance to a given query string.
Indexing structures for similarity search make use of the triangle inequality
to prune the search space. Consider an element p with an assosciated subset of
elements X such that \\
$\forall x \in X, d(p, x) <= k$ \\

We want to find all strings within edit distance e from given query string q.
That is reject all strings x such that 
\begin{equation}
d(q, x) > e
\label{cond1}
\end{equation}
From the triangle
inequality, $d(q, p) \leq d(q, x) + d(x, p)$. Hence $d(q, x) \geq d(q, p) -
d(x, p)$ which reduces to 
\begin{equation}
d(q, x) \geq d(q, p) - k 
\label{cond2}
\end{equation}
From equations (\ref{cond1}) and (\ref{cond2}), the criterion reduces to 
\begin{equation}
d(q, p) - k > e
\label{criterion}
\end{equation}
If the inequality is satisfied, the entire subset X is eliminated from
consideration. \\ 
However, we need to compute the O(mn) edit distance for all the elements in the
subsets that do not satisfy the above criterion. \cite{bagdist} proposes bag
distance which is given as 
\begin{equation}
bag\_distance = max(|x-y|, |y-x|)
\label{bag_dist}
\end{equation}
where $(x-y)$ is the set of the characters in x after dropping all common
elements and $|x-y|$ gives the number of characters in (x-y). The algorithmic
complexity for this computation is O(m+n) where $|x|$ = m, $|y|$ = n. Since
$d_{bag}(x, y) \leq d_{edit}(x, y)$, bag distance can be used to filter out
some of the candidate strings thereby reducing the search cost. 
\section{Index Structures}
In this section, we provide a brief description of the data structures used for
similarity indexing. Here,
\begin{itemize}
\item {U is the set of all strings.}
\item {n is the number of tuples in the dataset.}
\item {B is the bucket size, i.e., the maximum number of tuples a leaf node can hold.}
\item {d(a, b) is the edit distance between strings a and b.}
\item {q is the query string.}
\item {e is the {\it search distance}, i.e., all strings within an edit distance of e
from q should be returned on a proximity search.}   
\end{itemize}
\subsection{BK Tree}
The Burkhard-Keller tree(BK tree) presented in \cite{bk} is probably the first
general solution to search in metric spaces. A pivot element p is selected from
the data set U and the dataset is partitioned into subsets $U_i$ such that
($\forall u \in U_i, d(p, u) = i$). Each of the subsets is recursively partioned
until there are no more than B elements in a subset.\\
For a given query and search distance, the search starts at the root(pivot 
element p) and traverses all subtrees at distance i such that  
\begin{equation}
 d(p, q) - e \leq i \leq d(p, q) + e
\label{bk-test}
\end{equation}
holds and proceed recursively till a leaf node is reached. In the leaf node, the
query string is compared with all the elements.
\subsection{FQ Tree}
Fixed Queries trees \cite{fq} is a variation of BK trees. This tree is
basically a BK tree where all the pivot elements at the same level are
identical. The search algorithm is identical to that for BK trees. The benefit
of FQ trees over BK trees is that some of the comparisons between the query
string and the internal node pivots are saved along the backtracking that
occurs in the tree. 
\subsection{FH Tree}
In Fixed Height FQ trees \cite{fq}, all leaves are at the same height. This
makes some leaves deeper than necessary, but no additional costs are incurred
as the comparison between the query and intermediate level pivot may already
have been performed.   
\subsection{Bisector Tree}
Bisector tree(BS tree) \cite{bst} is a binary tree built recursively as
follows: Two routing objects $p_1$ and $p_2$ are chosen. While insertion, elements
closer to $p_1$ are inserted in the left subtree and those closer to $p_2$ are
inserted in the right subtree. For each routing object, the maximum covering
radius($r_i$), i.e., the maximum distance of $p_i$ with any element in its subtree is
stored. In our implementation, the distance of the element with its parent
routing object is also stored. This helps in reducing some of the distance
computations as shown in \cite{mtree}.\\
For a given query and edit distance, search starts at the root and recursively traverses
the left subtree if 
\begin{equation}
 d(p_1, q) - e \leq r_1 
\label{bst-test}
\end{equation}
and the right subtree if a similar condition holds for $p_2$.
\subsection{M Tree}
The bisector tree can be extended to m-ary tree \cite{mtree} by using m routing
objects in
the internal node instead of two. We select m routing objects for the first level.
Together with each routing object is stored a covering radius that is the maximum
distance of any object in the subtree associated with the routing object. A new
element is compared against the m routing objects and inserted into the {\it best
subtree} defined as that causing the subtree covering radius to expand less and
in the case of ties selecting the closest representative. 
Thus it can be viewed that associated with each routing object $p_i$, is a region of the metric space Reg($p_i) = (u \in U | d(p_i, u) < r_i$)
where $r_i$ is the covering radius. Further, each subtree is partitioned recursively. \\
In the internal node, $p_i$ and $r_i$ are stored together with a pointer
to the associated subtree. Further to reduce distance computations M tree also
stored precomputed  distances between each routing object and its parent.\\ For
a given query string and search distance, the search algorithm starts at the
root node and recursively traverses all the paths for which the associated
routing objects satisfy the following inequalities. 

\begin{equation}
 |d(p^p_i, q) - d(p^p_i,p_i)| \leq r_i + e 
\label{mtree-1}
\end{equation}
and
\begin{equation}
 d(p_i, q)  \leq r_i + e 
\label{mtree-2}
\end{equation}

In equation (\ref{mtree-1}), we take advantage of the precomputed distance between
the routing object and its parent.

\subsection{VP Tree}
Vantage Point tree(VP tree) \cite{vp} is basically a binary tree in which pivot
elements called {\it vantage points} partition the data space into spherical cuts at
each level to enable effective filtering in similarity search queries. It is built
using a top down approach and proceeds as follows. A vantage point $S_v$ is chosen from the dataset and the
distances between the vantage point and the elements in its subtree are
computed. The elements are then grouped into the left and right subtrees based
on the median of the distances, i.e., those elements whose distance from the
vantage point is less than or equal to the median is inserted in the left
subtree and others are inserted in the right subtree. This partitioning
continues till the elements in the subtree fit in a leaf. The median value M is
retained at each internal node to aid in the insertion and search process. In
addition, each element in both the internal and leaf node holds the distance entries for every
ancestor, which helps in cutting down the number of distance computations at
query time. An optimized tree can be obtained by using heuristics to select
better vantage points. \\
Search for a given query string starts at the root node. The distance between q
and the vantage point at the node($S_v$) is computed and left subtree is
recursively traversed if 
\begin{equation}
 d(q, S_v) - e \leq M 
\label{vp_cond1}
\end{equation}
Similarly, right subtree is traversed recursively if the following inequality
holds. 
\begin{equation}
 d(q, S_v) + e \geq M 
\label{vp_cond2}
\end{equation}
Once a leaf node is reached, the query string need to be compared with all the elements
in the leaf node, but some of the distance computations can be saved using the ancestral distance information. 
\subsection{MVP Tree}
VP tree can be easily generalized to a multiway tree structure called Multiple
Vantage Point tree \cite{mvptree}. A notable feature of MVP tree is that
multiple vantage points can be chosen at each internal node and each of them can partition the data space into m groups. Hence it is required to store multiple cut off values instead of a
single median value at each internal node. The various parameters that can be tuned to improve the
efficiency of MVP tree are 
\begin{itemize}
\item{the number of vantage points at each internal node (v).}
\item{the number of partitions created by each vantage point (m).}
\item{the number of ancestral distances associated with each element in the
leaf (p).}
\end{itemize}
The insertion procedure starts by selecting a vantage point $S_{v1}$ from the
dataset. The elements under the subtree of $S_{v1}$ are ordered with respect to
their distances from $S_{v1}$ and
partitioned into m groups. The m-1 cut off values are recorded at the internal
node. The next vantage point $S_{v2}$ is a data point in the rightmost
(m-1) partitions, which is farthest from $S_{v1}$ and it divides each of the m
partitions into m subgroups. It can be observed that the nth vantage point is
selected from the rightmost (m-n+1)
partitions and the fan out at each internal node is $m^v$. This is continued
until all elements in the subgroup fit in a leaf node. At the leaf, each element keeps
information about its distance from its first p ancestors. \\

Given a query string q and an edit distance e, q  is compared with the v vantage
points at each internal node starting at the root. Let the distance between the
vantage point $S_{vi}$ and q be d($S_{vi}$, q) and $M_i$ be the cut off value
between subtrees $T_i$ and $T_{i+1}$. $T_i$ is recursively traversed if the
both the 
inequalities 
\begin{equation}
d(S_{vi}, q) - e \leq M_i 
\label{mvp_cond1}
\end{equation}
and 
\begin{equation}
d(S_{vi}, q) + e \geq M_{i-1}  
\label{mvp_cond2}
\end{equation}
hold. For traversing the first subtree, only (\ref{mvp_cond1}) need to be satisfied.
Similarly, the inequality (\ref{mvp_cond2}) is used to traverse the last subtree. A detailed description of the search procedure can be
found in \cite{mvptree}.

\subsection{Clustering}
Another technique used in similarity searching to reduce search cost is
Clustering. Clustering partitions the collection of data elements into groups called
clusters such that
similar entities fall into the same group. Similarity is measured using the
distance function, which satisfies the triangle inequality. A representative
called clusteroid is chosen from each cluster. While searching, the query
string is compared against the clusteroid and the associated cluster can be
eliminated from consideration in case criterion (\ref{criterion}) does not
hold, which helps in reducing the search cost. \\
\cite{cluster} proposes BUBBLE for clustering data sets in arbitrary metric
spaces. The two distance measures used in the algorithm are given as \\ \\
{\bf RowSum} Let O = {$O_1, O_2, ..., O_n$} be a set of data elementsin
metric space with distance function d. The rowsum of an object o $\in$
O is defined as RowSum(o) = $\Sigma_{j=1}^n d^2(O, O_j)$. The clusteroid C is
defined as the object C $\in$ O such that $\forall o \in O : RowSum(C) \leq
RowSum(o)$.\\ \\ 
{\bf Average Inter-Cluster Distance} Let $O_1 = \{O_{11}, ..., O_{1n_1}\}$ and
$O_2 = \{O_{21}, ..., O_{2n_2}\}$ be two clusters with
number of elements n1 and n2 respectively. The average inter-cluster distance
is defined as $D_2 = \frac {\Sigma_{i=1}^{n_1} \Sigma_{j=1}^{n_2} d^2(O_{1i},
O_{2j})}{n_1n_2}^{\frac{1}{2}}$.\\ \\
Insertion in BUBBLE starts by creating a CF* tree, which is a height
balanced tree. Each non-leaf node has entries of the form ($CF^*_i$, $child_i$)
where $CF^*_i$ is the {\it cluster feature}, i.e., the summarized
representation of the subtree pointed to by $child_i$. The leaf node entries
are of the form ($CF^*_i$, $cluster_i$) where $CF^*_i$ is the clusteroid and
$cluster_i$ points to the associated cluster. When an element x is to be
inserted, it is compared against all the CF* entries in the internal node
using the average inter-cluster distance $D_2$ and the child pointer
associated with the closest CF* entry is followed. On reaching a leaf node, the cluster closest to x is the one having minimum
RowSum value. If the distance between x and the closest clusteroid is less
than a threshold value T, it is inserted in that cluster, a new clusteroid
is selected and the CF* entries in the path from root to this leaf node are
updated. In case the difference is greater than T, a new cluster is
formed. In our implementation, each element entry in the cluster contains its
distance with the clusteroid to reduce the number of distance computations.\\
For a given query string and search distance, the query is compared with all
the clusteroids. If it does not satisfy the (\ref{criterion}), the cluster
elements need to be searched for similar strings. The precomputed distances can be used to
eliminate some distance computations. 
\begin{figure*}[t]
\centering
\includegraphics[angle=270,width=0.5\textwidth]{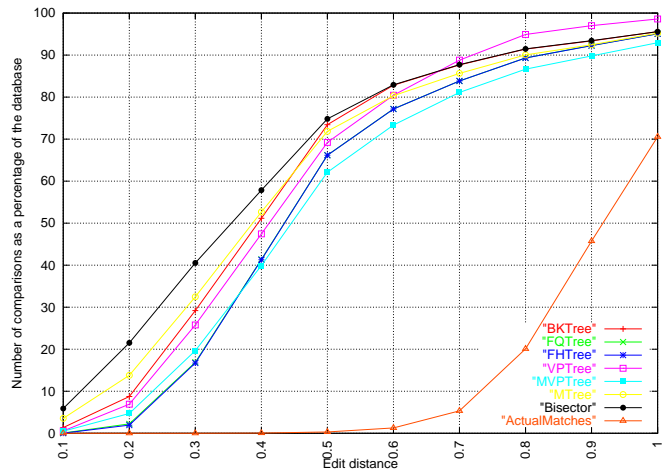}
\caption{Performance Comparison of Similarity Indexing Structures}
\end{figure*}
\subsection{MTB}
In case of M tree, a new element x is compared with the routing objects at the internal node and inserted into 
the best subtree. The best subtree is defined as the one for which the
insertion of this element causes the least increase in the covering radius of
the associated routing object. In the case of ties, the closest
representative is selected. This continues until we reach a leaf node. This may cause physically close elements to fall into different subtrees. 
Along the path, the covering radii of the selected routing objects are updated if x is
farther from p than any other element in its subtree. Thus there are no bounds
on the covering radii associated with the routing objects. A possible
optimzation is to bound the elements in the leaf nodes to be within a given
THRESHOLD of the routing objects in its parent node. Also, the new element is
inserted into the subtree associated with the closest routing object, there by
keeping the physically close elements together. These two optimizations result
in a new indexing structure, which we call {\it M Tree with Bounds}(MTB). Thus,
in the case of MTB the insertion of an
element that causes the covering radius of the routing object of the lowest
level internal to
exceed the THRESHOLD results in a partition of the leaf node entries such that
the THRESHOLD requirements are maintained. Searching is similar to that of the basic tree
implementation. 
\section{Experimental Setup}
We have performed an analysis of the various similarity indexing structures
described in the previous section. The metric used for comparing the
performance is the percentage of the database scanned for a given query and
search distance, which is a measure of the CPU cost incurred. \\
The experimental analysis were performed on a Pentium III(Coppermine) 768 MHz
Celeron machine running Linux 2.4.18-14 with
512 MB RAM. All the indexing structures were implemented in C. The O(mn) dynamic
programming algorithm to compute the edit distance between a pair of strings was used in the experiments. The dataset
used for the analysis was an English dictionary dataset comprising of 100,000 words.
The average word length of the dataset is around 9 characters. Six query sets
each of 500 entries were chosen at random from the data set for the
experiments. The results presented are obtained by averaging over the results
for these query sets. The page size is assumed to be 4K bytes. 
\section{Analysis}
In this section,we provide the analysis and the experimental results on the
performance of the various similarity indexing structures. The implementation details of the various index
structures are presented in the next subsection followed by the results.   
\subsection{Implementation Details}
Assuming a page size of 4K bytes, the bucket size is taken to be 512 entries
for BK tree, FQ tree and FH tree as each entry is 8 bytes. The routing data
elements are chosen at random from the dataset. \\
The leaf node for VP tree as proposed in \cite{vp} has a single entry. The
routing element is selected using the best spread heuristic \cite{vp}.  For
MVP trees, we ran the experiments for different values of parameters m, v and
p and the values 2, 2 and 10 were shown to give better performance. For p =
10, the number of leaf node entries is found to be 110. The vantage point is
selected at random for MVP tree. \\
In the case of bisector tree and M tree, the two farthest elements are chosen as pivot
elements at the time of split of a FULL node. For M tree, we ran the experiment
with the number of entries in the internal node m taking values 5 and 254. \\
In Clustering and Indexing with bounds, the THRESHOLD value used in our runs
was chosen to be 5. 
\subsection{Experimental Results}
\subsubsection{Search complexity}
In all the indexing structures, the criterion (\ref{criterion}), which is 
obtained from the triangle inequality is used to prune the search space. As
the search distance is increased, the number of pivots(or routing objects)
that fail to satisfy the criterion (\ref{criterion}) also increases resulting
in an increase in the percentage of the database scanned.\\

Figure 1 shows the performance of the various similarity indexing structures
with variation in the search distance. It can be seen that FQ tree and FH tree
perform better than BK tree. This can be attributed to two reasons: 
The number of pivot element comparisons is less in case of FQ tree and FH tree
as these trees have one fixed key per level. Whereas, in case of BK tree,
there are as many distinct pivot elements per level as the number of nodes at
that level. 
Further, FQ tree and FH tree use a better splitting technique resulting in
more partitions as compared to BK tree. Hence, some of the partitions can be
eliminated using (\ref{criterion}). 
\begin{figure}[t]
\centering
\includegraphics[width=0.4\textwidth]{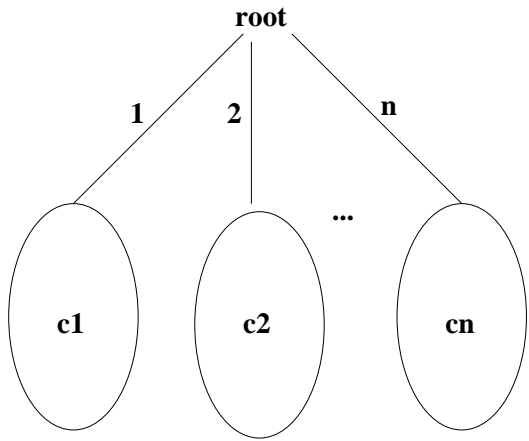}
\caption{Splitting characteristic of BKTree}
\end{figure}
Consider the case when a subset $C_i$ as shown in figure 2 is to be split in BK tree. Then the pivot
 element selected is some $c \in C_i$. Thus the maximum number of partitions
 that can result is 2i. However, in case of FQ tree, since a fixed pivot
 element is selected for each level, the chosen pivot is away from the subset,
 which may result in more partitions.
 It is shown in \cite{vp} that this results in better performance.
 In FH tree all the leaves are at the same level. Also, since we have already
 performed the comparison between the query and pivot of an intermediate level
 , we eliminate for free the need to consider a leaf. Hence FH tree performs
 slightly better than FQ tree.\\

Our implementation of VP tree uses the best spread heuristic \cite{vp} for
 selecting the vantage points. In addition, each internal node maintains the
 lower and upper bounds of the distance of elements in left and right
 subtrees. This can be used to cut down the distance computations using the
 triangle inequality. Because of these factors the performance is better as
 compared to BK tree. However, just like BK tree, as the vantage point is selected
 from the subset that is being partitioned and there are multiple distinct
 vantage points at any given level, FQ tree and FH tree show better
 performance. \\

As can be seen from the plots in Figure 1, MVP tree outperforms VP tree. Each leaf node entry in the MVP tree stores its distance to the first 10
 ancestors. These precomputed distances help in reducing the search cost as
 compared to VP tree. In addition, MVP tree needs two vantage points to
 partition the data space into four regions whereas VP tree requires three
 vantage points for the same. This further reduces the number of distance
 computations at the internal nodes at search time. The left partition
 obtained using vantage point $S_{v1}$ is partitioned again using       the
 farthest point $S_{v2}$ which is present in the right partition. 
 Also,  for smaller values of the edit distances($\leq 0.4 $) the internal node
 comparisons dominate the results. In case of the MVP tree, since there are
 multiple keys at each internal node, it results in more distance computations
 as compared to the FH and FQ tree, which have one fixed key per level. This explains the crossover in the
 curves of the FQ tree, FH tree and MVP tree at search distance 0.4.\\
\begin{figure*}[t]
\centering
\includegraphics[angle=270,width=0.5\textwidth]{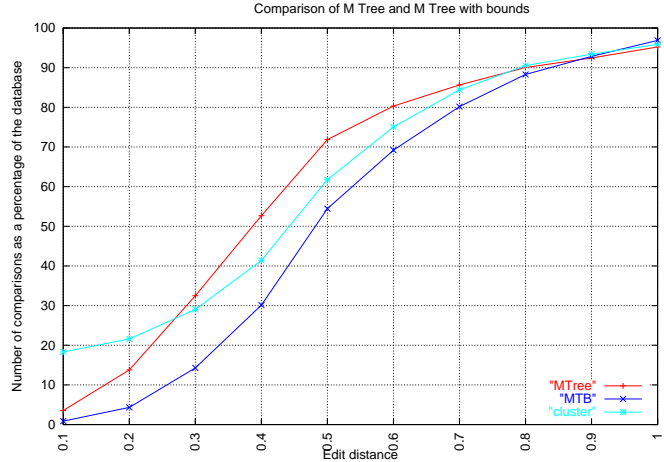}
\caption{Performance Comparison of M Tree, Clustering and MTB}
\end{figure*}

The clustering technique partitions the dataset into a fixed number
 of clusters $N_c$. This number varies inversely as the THRESHOLD i.e. the
 cluster radius. At search time, the query string is compared against each
 of the cluster representatives, the clusteroids. These comparisons are
 performed irrespective of the search distance. For a THRESHOLD of five, the
 clustering algorithm partitioned the dataset into 17,912 clusters. This
 explains the comparitively large number of searches for smaller values of
 search radii in figure 1. For clusteroids that do not
 satisfy the test in (\ref{criterion}), the associated cluster elements are
 sequentially compared against the query string.\\
 
 In the case of bisector tree, the insertion of a new data element may result
 in an increase in the covering radius of the routing object. The covering
 radii values depend upon
 the order in which the elements are inserted and may have large values.Due to
 this, at search time, a number of routing objects satisfy the test in equation
 (\ref{mtree-1}). Thus, the Bisector Tree shows poor performance as compared to
 the other indexing structures. With M trees, even though the new element is
 inserted into a subtree such that the resulting increase in the covering
 radius is the least, there are no bounds on covering radius value. So the
 performance is identical to that of bisector tree. The poor performance can
 also be attributed to the presence of more number of routing objects to
 partition the data space.\\

It can be observed from the graph in Figure 3 that MTB that combines the
features of M tree and clustering shows better performance.
 This can be  attributed to the two optimzations used, which result in well
 formed clusters. For lower values of the search distance, the overhead of the
 comparisons with large number of routing objects at the internal nodes results
 in poor performance. \\
\begin{figure}[h]
\centering
\includegraphics[angle=270,width=0.5\textwidth]{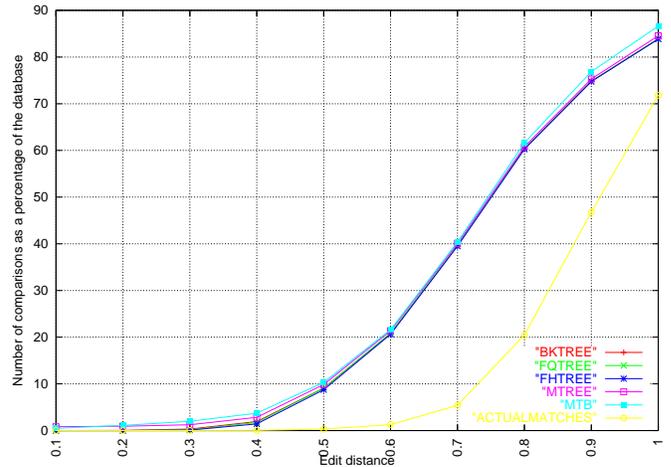}
\caption{Performance Comparison of Indexing Structures using bag distance }
\end{figure}
 The graph in Figure 4 shows comparison of the various indexing structures when
 bag distance computation is used to reduce some of the edit distance
 computations. The graph shows the edit distance computations needed with search distances 
 varying from 10 to 100\%.

\begin{table}
\begin{center}
\begin{tabular} {|l|c|}
\hline
{\em Index Structure} & {\em Search Time (ms)}
\\ \hline
BK tree	&  0.5789	\\
BK tree (with bag distance) & 0.4164	\\
FQ tree & 0..5825 \\
FQ tree (with bag distance) & 0.4124 \\
FH tree & 0.5746 \\
FH tree (with bag distance) & 0.4090 \\
VP tree & 0.4951\\
M tree (with bag distance) & 0.3041\\
Cluster & 0.6531	\\
MTB (with bag distance)  & 0.1465 \\
\hline
\end{tabular}
\end{center}
\caption{Time complexity}
\label{tab:mytable}
\end{table}

\subsubsection{Search Time }
Table 1 lists the average search time(ms) per query taken by various indexing
structures. It can be observed that MTB tree takes comparatively lesser time. Bag distance computation helps in reducing the
time complexity.  
\section{Conclusions and Future Work}
We have presented a performance study of the search efficiency of similarity
indexing structures.  MTB, which combines the features of
clustering and M tree is found to perform better than all the other
indexing structures for most search distances. Bag distance
computation,
which is cheaper than edit distance computation, was used in the experiments.
Its use resulted in reduced time complexity. Further, in applications where the
required search distance is low and the string lengths are large, even better
performance might result.\\

It can be observed that index structures like MVP tree, which make use of
precomputed distances with ancestors
to prune the search space perform better than others. In similarity
searching, since multiple paths are traversed, keeping a fixed key per level
as in FQ tree minimizes the search cost by reusing the precomputed distance
at that level. Thus, reusing the pre computed distances results in better
performance. Some indexed structures were shown to perform better with
smaller 
values of edit distances($e \leq 0.3$) whereas some others perform better at
higher values. 
It would be advantageous to maintain multiple index structures and depending upon the edit distance, select the appropriate one. 
Using cheaper distance computation algorithms can also significantly reduce the CPU
cost. The quality of partioning is largely dependent on the
heuristic used for selecting the pivots. As a future work, we propose to
analyse the performance of various index structures with different heuristics.  
\section{Acknowledgement}
We thank A Kumaran of Database Systems Lab, IISc for his advice during the work.
\bibliography{}

\end{document}